\documentclass[a4paper,11pt]{article}
\usepackage{jheppub} % for details on the use of the package, please see the JINST-author-manual
\usepackage{float}      % for [H]
\usepackage{placeins}   % for \FloatBarrier
\usepackage{lineno}
\usepackage[dvipsnames]{xcolor}
\newcommand{\hmc}{\mbox{HMC}}
\newcommand{\hmcc}{\mbox{HMC}_C}
\title{\boldmath Eigenvalue-cluster Algorithm for Matrix Monte Carlo}

\author{S. Kováčik and M. Hrmo}
\affiliation{Comenius University,\\
Bratislava, Slovakia}

\emailAdd{samuel.kovacik@fmph.uniba.sk}

\abstract{Various physical models can be expressed in terms of matrices. A valuable tool for analysing matrix models is numerical simulations, often the Metropolis algorithm with various improvements. The downside of this approach is that the simulation may become stuck in a vacuous state, and probing the relevant parts of the configuration space might be difficult. Here, we propose an algorithm that moves around a cluster of eigenvalues and show that it converges to the true vacuum state. }

\begin{document}
\maketitle
\flushbottom

\section{Introduction}
Matrix models are an important part of modern physics. They appear in various forms in M-theory \cite{Banks:1996vh,Hanada:2013rga}, serve as effective models of quantum space \cite{Ydri:2014rea, Barrett:2015foa}, but also find application in other areas \cite{Guhr:1997ve,Shimamune:1981qf,Barrett:2015foa}. There are various methods of analysing such models; sometimes direct calculations are possible, but often these are prohibited, and one can access the features of the model using semi-analytical approximations \cite{Tekel:2015zga}, a bootstrap algorithm \cite{Lin:2020mme} or Hamiltonian Monte Carlo (HMC) simulations \cite{Jha:2021exo}. In this paper, we will focus on the numerical analysis of matrix models. 

One of the most important descriptions of a model is its phase diagram --- that is, the identification of the model's phases and their boundaries. Here, matrix models contain new phases that might not be present in continuum theories, which are conjectured to be described in the limit of infinite matrix size. Let us mention a well-known example, the scalar field theory on the fuzzy sphere. Simply, this model can be expressed as a matrix model, and the phase is characterised by the nature of the eigenvalue distribution. The model on an ordinary sphere has two distinct phases: one characterised by $\left\langle \phi \right\rangle = 0$ and the other by $\left\langle \phi \right\rangle \neq 0$. The theory with the same naive action on the fuzzy sphere has three phases. There is one phase for which all of the eigenvalues are close to zero. Then, there is a phase in which all of the eigenvalues are close to a constant value $\phi_0\neq 0$. Both phases have counterparts in the ordinary sphere. However, there is an additional phase in which half of the eigenvalues are close to $\phi_0\neq 0$ and the other half is close to $-\phi_0$, for more details see \cite{Kovacik:2018thy}.

With this vacuum structure, numerical simulations may be difficult, as local free-energy minima may prevent the system from moving to the global minimum. This means that if one initiates the simulation in a phase with all eigenvalues close to a single value and the global minimum corresponds to a two-cut solution around $\pm \phi_0$, the eigenvalues have to tunnel through a high potential barrier. In \cite{Kovacik:2022kfh}, the authors have suggested an algorithm to overcome this issue called the eigenvalue-flipping algorithm. The idea is that during the Metropolis algorithm, one first does the eigenvalue decomposition of the current (matrix) state, flips some of the eigenvalues and recomposes the matrix. As the potential terms are not affected by this procedure (the kinetic term is), there is a good chance it will be accepted during the procedure. 

This method works well for models that have potentials with the $\phi \rightarrow - \phi$ symmetry. However, many models lack it. For example, attention has been recently drawn to a class of Dirac ensemble models \cite{Hessam:2021byc}. It has been shown that, for some parameters, these models exhibit a rich set of local minima. In those cases, the eigenvalues are inequivalently split and are close to two distinct values, $\phi_1, \phi_2$, with $|\phi_1| \neq |\phi_2|$. Unless the simulation is initiated near the global minimum of the effective action, it might not be able to cross the potential barrier and reach ergodicity. For models with such a rich (false) vacuum structure, performing a successful ordinary Metropolis algorithm might not be feasible. 

In this paper, we propose and analyse an algorithm to overcome this issue. The main idea is that we first perform the standard eigenvalue decomposition, but then, instead of just flipping the signs of some of the eigenvalues, a cluster of eigenvalues is selected and moved \textit{en masse}. We show that, for the considered class of Dirac ensemble models and other examples, this method performs very well at probing the configuration space and reaching the global minimum of the potential. This model was chosen as a particular example, but the method does not rely on its details and should work for a large class of models with similarly rich structure.

\section{Eigenvalue probability distribution}

Matrix models are defined by two objects: first, a set of matrices, for example, a set of Hermitian matrices of given fixed size $N$. In addition to this, one needs to provide a probabilistic measure. This is typically done by specifying $S(\phi)$, the action, and then the probability is proportional to
\begin{equation} \label{prob}
    p(\Phi) = Z^{-1} e^{- S(\Phi)},
\end{equation}
where $Z$ is the partition function. From this, one can define and compute various observables 
\begin{equation}
    \left\langle {\cal{O}} \right\rangle = Z^{-1} \int {\cal{O}} e^{- S(\phi)} d [\Phi],
\end{equation}
here $d [\Phi]$ denotes integration over all possible matrix elements. 

In models, the action takes many different forms. It can be a polynomial in $\phi$, or joined with other objects, for example, arising from an underlying geometry of the Laplace operator \cite{Tekel:2015zga} or from the curvature of the space as in the Grosse--Wulkenhaar model \cite{Prekrat:2022sir}. Some models contain only single-trace terms, others are called multi-trace models; the space of matrix models is vast. Models containing only a single-trace polynomials of $\phi$ are sometimes referred to as pure-potential models. In those, the action depends only on the eigenvalues of $\phi$ since $\mbox{Tr }(U^{-1} \Lambda  U) = \mbox{Tr } \Lambda$, where $\Lambda$ is a diagonal matrix. When performing the change of integration coordinates $\Phi \rightarrow \Lambda$ and integration over $U$, one effectively introduces a logarithmic repulsion term $\sim \sum \limits_{i\neq j} \log |\lambda_i - \lambda_j|$, where $\lambda_i$ are the elements of $\Lambda$. Therefore, even if the resulting action depends only on $N$ eigenvalues $\lambda$, the model still remembers being a matrix one. In practice, the behaviour of the eigenvalue system is captured by the free energy
\begin{equation} \label{freeEnergy}
A= S -\sum \limits_{i \neq j} \log |\lambda_i -\lambda_j|.
\end{equation}

In most models, integrating out $U$ is difficult, and one cannot simplify the model only to $S(\Lambda)$. However, analysing the behaviour of eigenvalues can still be valuable, and many models exhibit distinct phases that differ in the nature of their eigenvalue distribution; for example, their symmetry, support cuts, and so on. For some models, different phases coexist, meaning that there are multiple eigenvalue distributions that minimise the free energy, but only one is the global minimum. The eigenvalue distribution is sensibly defined only in the limit of infinite matrix size, but can be estimated from finite-matrix numerical simulations.

However, the numerical simulations of matrix models can be tedious. Consider a model that has, for a given set of parameters, two distinct local minima of the free energy, that is 
\begin{equation}
    A'(\rho_1) = A'(\rho_2) = 0.
\end{equation}
For example, $\rho_1$ can be a symmetric two-cut distribution and $\rho_2$ can be an asymmetric one-cut solution. In this case, a numerical simulation would start from an initial configuration, move toward one of the solutions (which one depends on the initial configuration), and, depending on the simulation's parameters, can become stuck in the vicinity of the wrong one. In \cite{Kovacik:2022kfh}, a solution was proposed that works well for models with $\Phi \rightarrow - \Phi$ symmetry. Simply, first the eigenvalue decomposition is made, $\Phi = U^{-1} \Lambda U$, then some randomly chosen eigenvalues have their signs flipped, forming $\Lambda'$, which is then remade back to $\Phi = U^{-1} \Lambda' U$, which is then proposed to the Metropolis algorithm check. This algorithm also worked well for some models in which this symmetry holds only for some terms in the action, for example, for scalar field theories on fuzzy spaces, where only the kinetic terms break it. 

An important property of the tested models that enabled this algorithm to work well was the large overlap between the supports of $\rho_1$ and $\rho_2$. For example, when there was one two-cut solution and one one-cut solution, one of the cuts overlapped significantly with the single cut of the other solution. That means, for example, that starting from an asymmetric one-cut solution, this algorithm allows the simulation to find the symmetric two-cut solution reasonably quickly (and vice versa).

However, this algorithm does not perform well for models with a different structure of (local) solutions. If, for example, there were two distinct two-cut solutions well separated from each other, or if there were a set of asymmetric solutions, this algorithm would not find them. 

Therefore, to numerically analyse a large class of matrix models with a rich solution structure, a novel algorithm is needed to efficiently probe the configuration space during simulation. In the next section, we introduce such an algorithm.

\section{Cluster algorithm}

The eigenvalue flipping algorithm used to perform eigenvalue decomposition of the matrix $\Phi$ begins with the decomposition
\begin{equation} \label{EVdecomp}
    \Phi = U^{-1}  \begin{pmatrix}
  \lambda_1 & & & \\
  & \lambda_2 & & \\
  & & \lambda_3 & \\
  & & & \ddots
\end{pmatrix}  U.
\end{equation}
Here, we assume the eigenvalues are ordered, that is, $\lambda_1>\lambda_2>...$. While the eigenvalue algorithm only flipped the signs of some eigenvalues, we propose a more delicate change that will scan the configuration space of a wide class of matrix models more efficiently. 

The basic idea of the algorithm, which is from now on denoted $\hmcc$, is that a cluster of eigenvalues will be shifted together. To do this, one first randomly selects an eigenvalue and then takes all others from its vicinity. All of them are then multiplied by the same factor $\alpha$, which in principle can be negative to include essential features of the eigenvalue flipping algorithm.

To be more precise:

\begin{enumerate}
    \item Save the old configuration $\Phi_i$.
    \item Perform the ordered eigenvalue decomposition \eqref{EVdecomp}.
    \item Choose random $1 \le i \le N$.
    \item Do $\lambda_j \rightarrow \alpha \lambda_j$ for all $\lambda_j$ such that $|\lambda_i - \lambda_j|<\beta$.
    \item Recompose the matrix, $\Phi_f$, using the same $U$ in \eqref{EVdecomp} and updated eigenvalues.
    \item Perform the Metropolis check as defined in \eqref{Metropolis}.
\end{enumerate}

Within our test we chose $\alpha$ and $\beta$ to be also randomly chosen parameters, in more detail
\begin{equation} \label{multiplication}
    \alpha = (-1)^\rho{\cal N} (1, \sigma_\alpha),
\end{equation}
where ${\cal N}$ denotes a random number from a normal distribution, where a natural choice is to take $\rho$ is $0$ with some probability $p$ and $1$ with the probability $1-p$. For the second parameter, we have

\begin{equation}
    \beta = {\cal N} (\mu_\beta, \sigma_\beta) \left(\lambda_1-\lambda_N\right),
\end{equation}
Basically, stating that a naturally chosen fraction of the whole distribution support (ignoring gaps) is picked. This leaves four parameters for the algorithm: $\sigma_\alpha, \mu_\beta, \sigma_\beta, p$. Their choices are optional, as are the other parameters of the HMC algorithm, and different choices suit different models. 

Given the nature of the proposed step, one must be careful with the Metropolis check to preserve the simulation's detailed balance after thermalisation. When making the proposal $\Theta_1\to\Theta_2$, the acceptance probability is, in general, given as
\begin{equation}
    P(\Theta_1\to\Theta_2)=\min \left(1, \frac{\rho(\Theta_2)q(\Theta_2\to\Theta_1)}{\rho(\Theta_1)q(\Theta_1\to\Theta_2)}\right),
\end{equation}
where $\rho(\Theta)=e^{-A(\Theta)}$ is the probability with which $\Theta$ should appear in the final ensemble and $q(\Theta_1\to\Theta_2)$ is the forward proposal probability, the probability with which $\Theta_2$ is proposed as a next step in the chain after $\Theta_1$. We call $q(\Theta_2\to\Theta_1)$ the reverse proposal probability. For a standard HMC step, the proposal probability is symmetric and thus $\frac{q(\Theta_2\to\Theta_1)}{q(\Theta_1\to\Theta_2)}=1$. Therefore, the Metropolis-Hastings acceptance probability for a $\hmc$ step is
\begin{equation}
    P_{\text{HMC}}(\Theta_1\to\Theta_2)=\min \left(1, e^{-(A(\Theta_2)-A(\Theta_1))}\right)=\min(1, e^{-\Delta A}).
\end{equation}
For the $\hmcc$, the proposal probability is no longer symmetric, $\frac{q(\Theta_2\to\Theta_1)}{q(\Theta_1\to\Theta_2)}\neq1$. This asymmetry comes in two parts. Firstly, suppose that the size of the rescaled cluster is $m$. This means that $m$ eigenvalues are scaled by $|\alpha|$ and possibly sign-flipped. The proposal probability for the sign flip is symmetric, so it is sufficient to consider the rescaling by $|\alpha|$. This transform in the eigenvalue space contributes to the proposal probabilities ratio as the (inverse value of) the Jacobian of the transform. The Jacobian is $J=|\alpha|^m$, therefore a factor of 
\begin{equation}
    J^{-1}=|\alpha|^{-m},
\end{equation} 
is needed in the Metropolis-Hastings acceptance probability.
Secondly, we must not forget that the inverse proposal would rescale the cluster by the multiplicative factor $1/\alpha$, which is not sampled with equal probability as $\alpha$. In the forward proposal, this is $\sim \exp{\left(-\frac{(\alpha-1)^2}{2\sigma_\alpha^2}\right)}$ for the reverse proposal we have to switch $\alpha\to1/\alpha$. This contributes to the Metropolis-Hastings acceptance probability by the factor of 
\begin{equation}
    q_{\alpha}=\exp{\left(-\frac{(|\alpha|^{-1}-1)^2-(|\alpha|-1)^2}{2\sigma_\alpha^2}\right)}.
\end{equation} 
Therefore, the total Metropolis-Hastings acceptance probability for the cluster-moving step is
\begin{equation} \label{Metropolis}
    P_{\mbox{\scriptsize{HMC}}_{\scriptsize{C}}}(\Lambda_1\to\Lambda_2)=\min\left(1,e^{-\Delta A}J^{-1}q_\alpha\right)=\min\left(1, e^{-\Delta A-\frac{(|\alpha|^{-1}-1)^2-(|\alpha|-1)^2}{2\sigma_\alpha^2}}|\alpha|^{-m}\right).
\end{equation}

We tested the $\hmcc$ algorithm on some well-known models: the fuzzy sphere scalar field theory, the Grosse-Wulkenhaar model, and the asymmetric multi-trace model; in each case, it outperformed the standard HMC. Details are in the appendices. We now move our attention to a model with a delicate vacuum structure. 

\section{Example: Dirac $(1,0)$ model}

Let us now consider a model of random Hermitian matrices of size $N$ in which the probability \eqref{prob} is given by the action

\begin{equation} \label{action}
 S =2N \left(\mbox{tr } \Phi^4 + g \mbox{ tr } \Phi^2 \right) + 2g \left(\mbox{tr }\Phi\right)^2 + 8\mbox{tr }\Phi \mbox{ tr }\Phi^3 + 6 \left(\mbox{tr }\Phi^2\right)^2,
\end{equation}
which depends on a single parameter $g$. This model might seem arbitrary, but we chose it because it has a rich vacuum structure for some parameter range, to be exact, for values $g<-3.187$, the model has various asymmetric solutions, \cite{DArcangelo:2026kka}. The other reason is that this model is one of the prominent examples of Dirac models, which has been studied thoroughly recently \cite{Barrett:2015foa}. 

We have performed a set of simulations. First, they differ by $N$, we picked values of $N=100,150,200$ as these are reasonably large for standard numerical simulations. For each value of $N$, we picked four values of $g$: two below and two above the critical value, and two reasonably close to it. For each such set of parameters, we performed the simulation without and with the cluster algorithm. We purposely ran very short HMC chains of $1000$ steps with an acceptance rate close to $ 100\%$, and we saved the eigenvalues of every tenth configuration. For each simulation, we show the full list of eigenvalues as a function of the simulation time $t_{\mbox{\scriptsize{HMC}}}$, the free energy, eigenvalues for the first $100$ HMC steps and the probability distribution of eigenvalues as obtained from the last $10$ HMC steps. It is expected that the two simulations shown on the left side will find the asymmetric two-cut phase, while those on the right will find the symmetric two-peak one-cut distribution. All simulations were initialised from a configuration $\Phi=\mbox{diag} \left(1,1,\dots \right)$ which is close to an asymmetric one-cut solution, which is known to exist in the model for some values of $g$ but is not the preferred solution. We initiated the system from this configuration to see if the simulations get stuck in the false vacuum state. 

\begin{figure}
    \centering
    \includegraphics[width=1\linewidth]{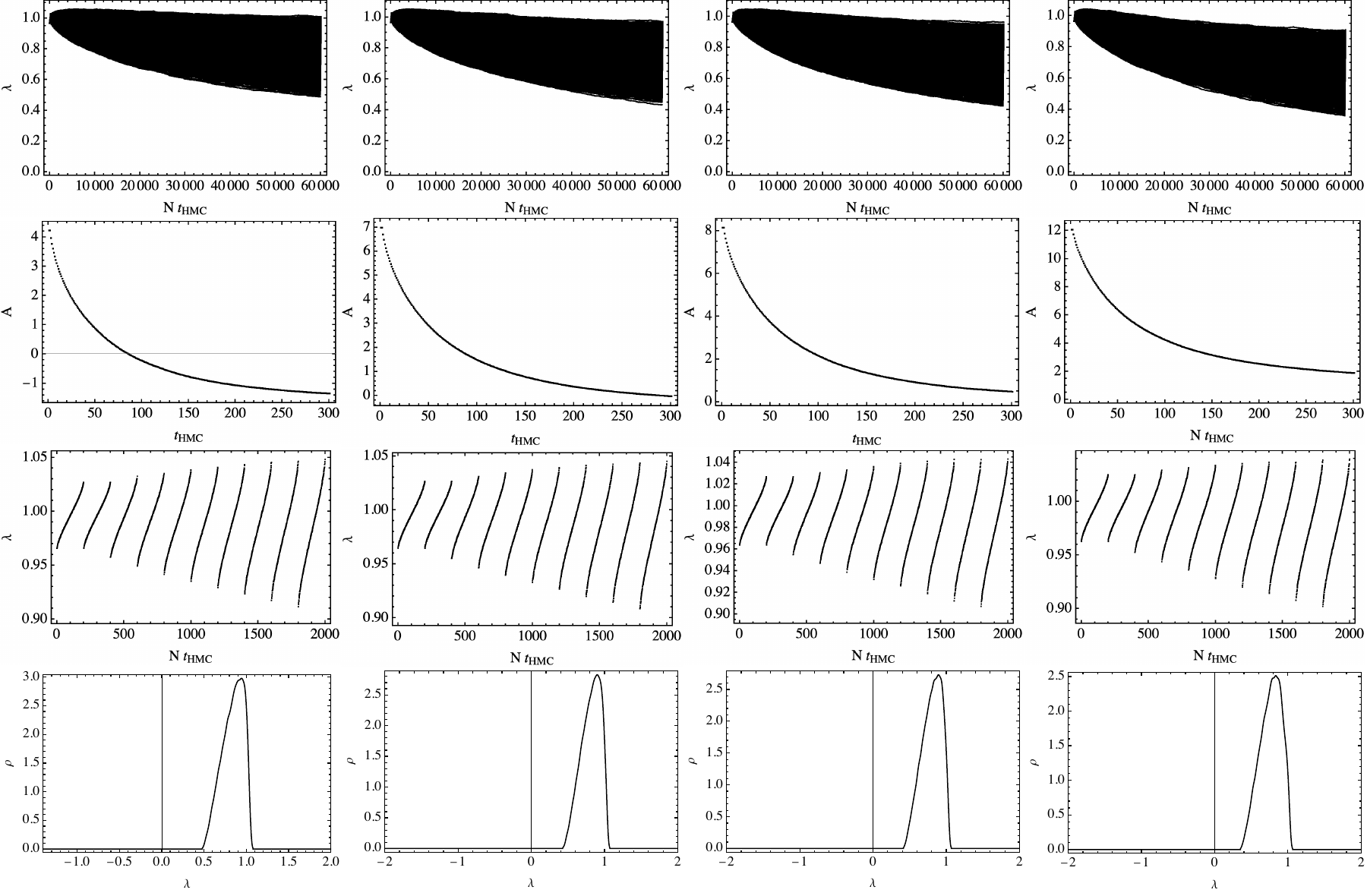}
    \caption{Results of the numerical simulations of the $N=200$ model specified in \eqref{action} with the standard HMC, columns are for $g=-4,-3.3,-3,-2$. The lines show: eigenvalues over simulation time, free energy over simulation time, eigenvalues over simulation time (zoomed on initial steps) and the final eigenvalue probability distribution.}
    \label{fig:200}
\end{figure}

\begin{figure}
    \centering
    \includegraphics[width=1\linewidth]{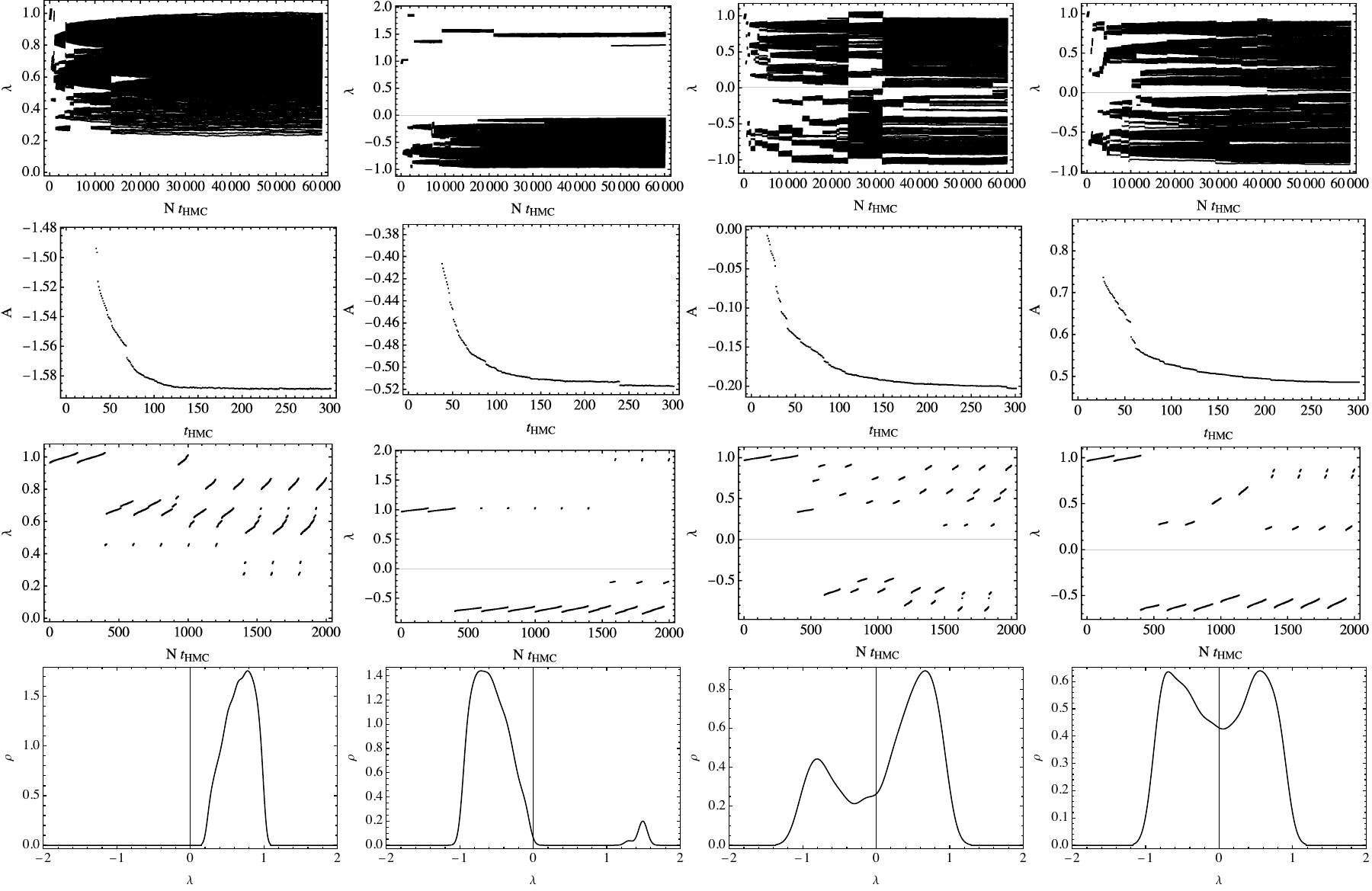}
    \caption{Results of the numerical simulations of the $N=200$ model specified in \eqref{action} with the cluster-algorithm improved HMC, columns are for $g=-4,-3.3,-3,-2$. The lines show: eigenvalues over simulation time, free energy over simulation time, eigenvalues over simulation time (zoomed on initial steps) and the final eigenvalue probability distribution.}
    \label{fig:200C}
\end{figure}

\begin{figure}
    \centering
    \includegraphics[width=1\linewidth]{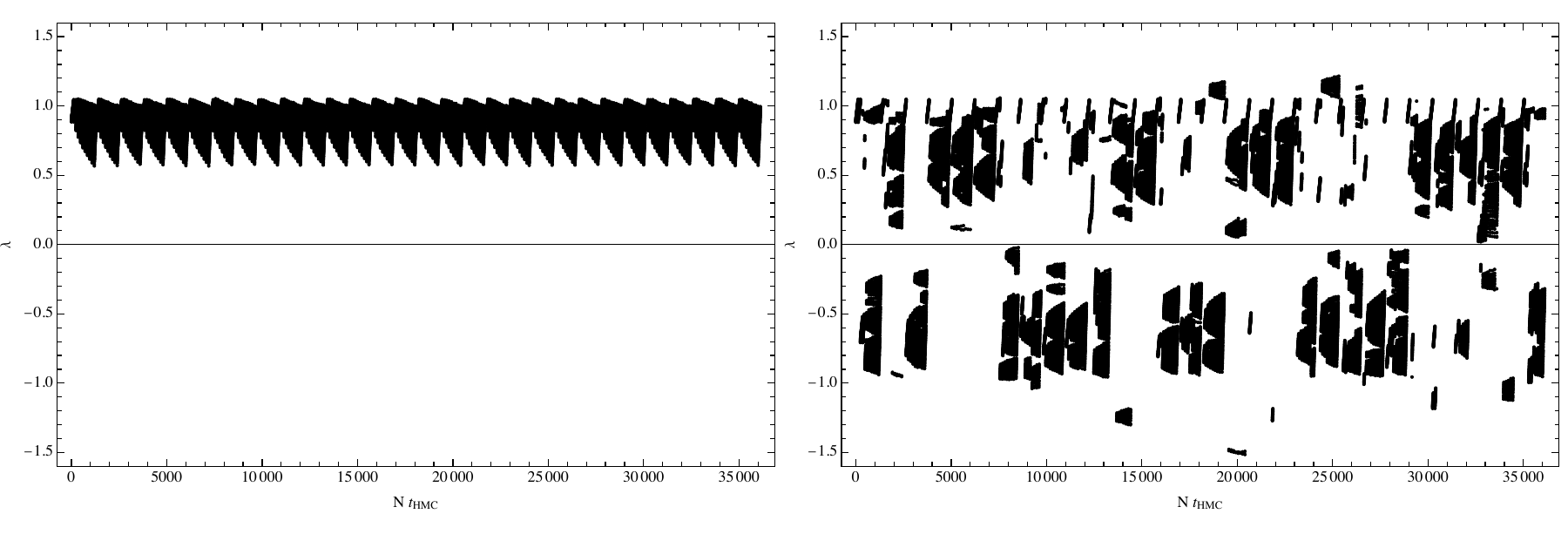}
    \caption{Comparison of the standard HMC (left) and cluster-algorithm enhanced HMC (right) for a simulation of the model \eqref{action} for $N=120, g=-3.3$ for which the system was restarted to the initial configuration after a fixed number of steps.}
    \label{fig:comp}
\end{figure}

\begin{figure}
    \centering
    \includegraphics[width=1\linewidth]{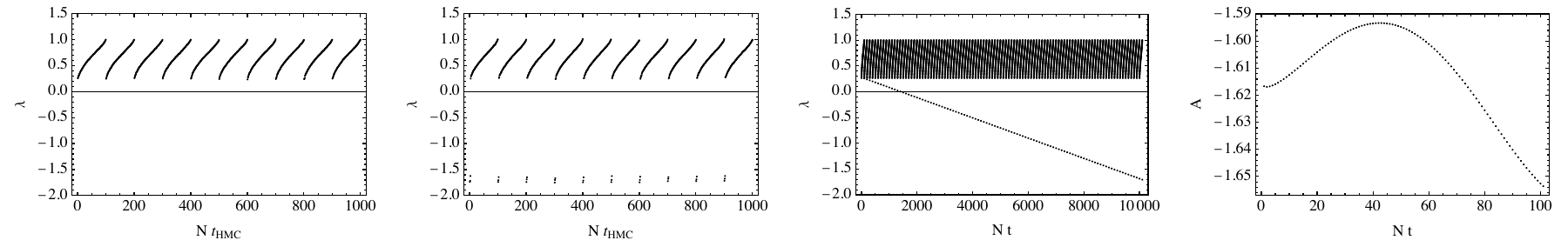}
    \caption{First two plots show the eigenvalues over the final steps of $N=100, g=-4$ simulation of the system \eqref{action} (standard and cluster-algorithm HMC). The third plot shows one eigenvalue manually shifted; the last plot shows the free energy of the consecutive configurations. }
    \label{fig:mig}
\end{figure}

\begin{figure}
    \centering
    \includegraphics[width=1\linewidth]{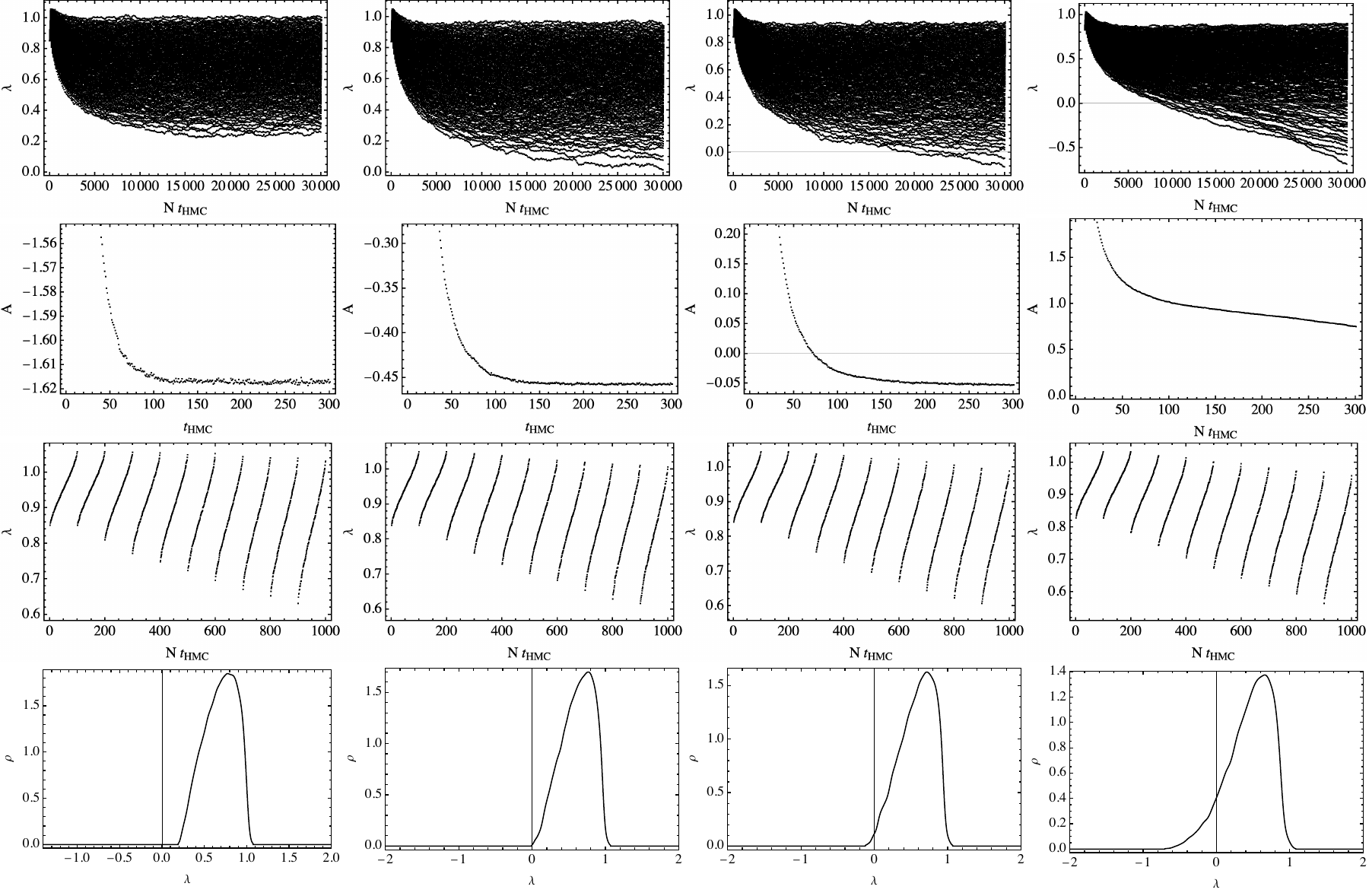}
    \caption{Results of the numerical simulations of the $N=100$ model specified in \eqref{action} with the standard HMC, columns are for $g=-4,-3.3,-3,-2$. The lines show: eigenvalues over simulation time; free energy over simulation time; eigenvalues over simulation time (zoomed on the initial steps); and the final eigenvalue probability distribution.}
    \label{fig:100}
\end{figure}

\begin{figure}
    \centering
    \includegraphics[width=1\linewidth]{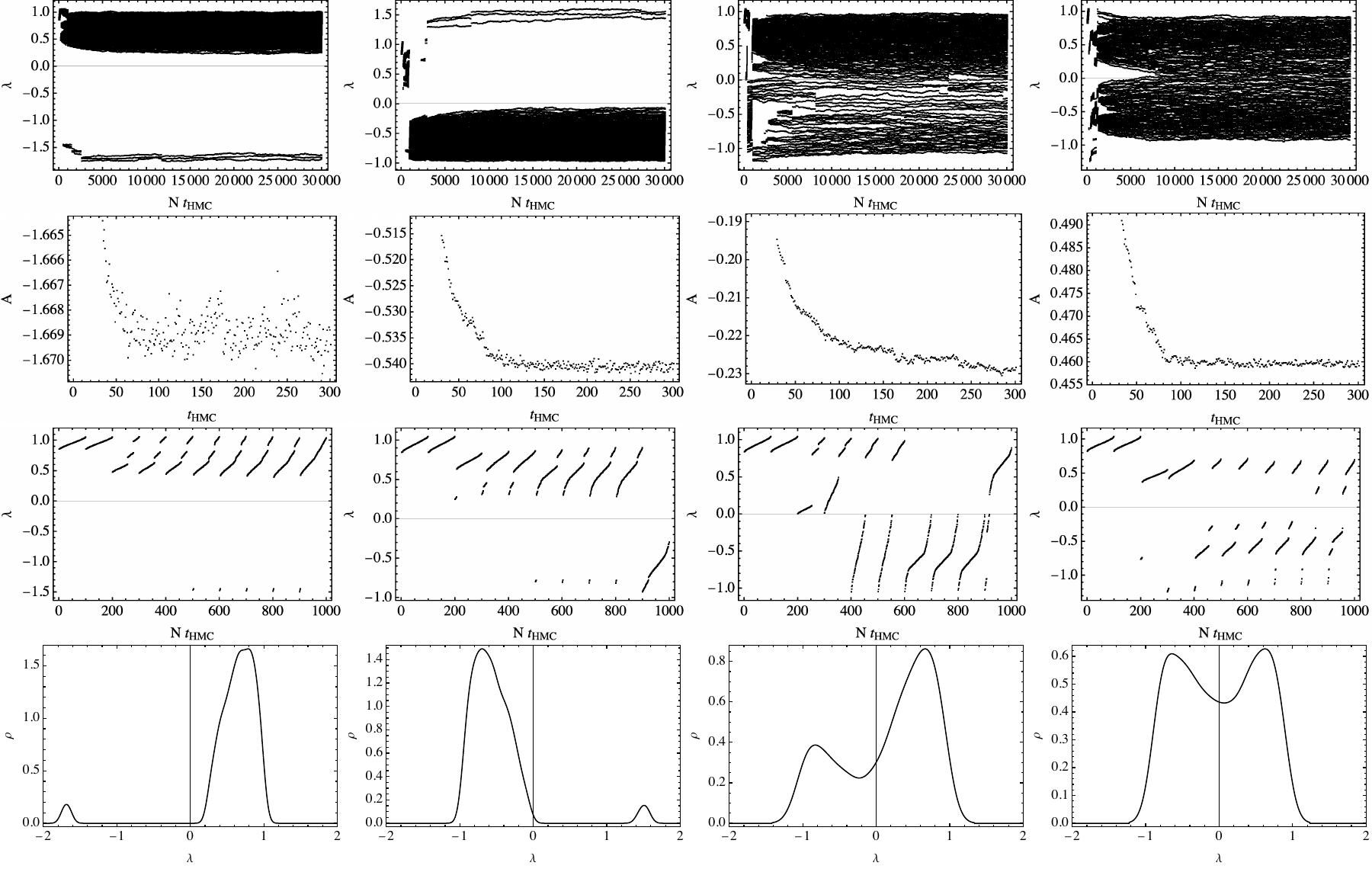}
    \caption{Results of the numerical simulations of the $N=100$ model specified in \eqref{action} with the cluster-algorithm improved HMC, columns are for $g=-4,-3.3,-3,-2$. The lines show: eigenvalues over simulation time; free energy over simulation time; eigenvalues over simulation time (zoomed on the initial steps); and the final eigenvalue probability distribution.}
    \label{fig:100C}
\end{figure}

\begin{figure}
    \centering
    \includegraphics[width=1\linewidth]{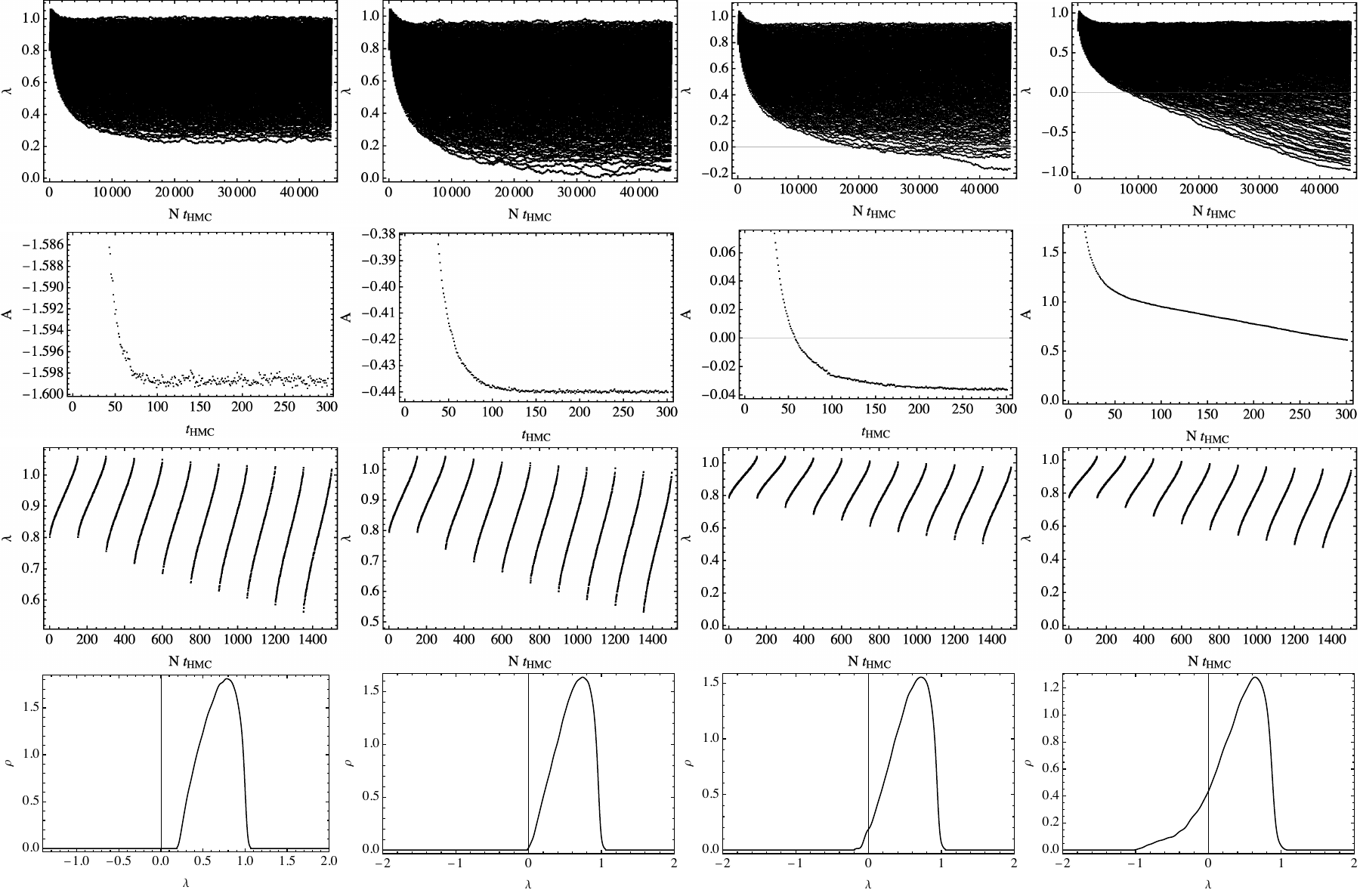}
    \caption{Results of the numerical simulations of the $N=150$ model specified in \eqref{action} with the standard HMC, columns are for $g=-4,-3.3,-3,-2$. The lines show: eigenvalues over simulation time; free energy over simulation time; eigenvalues over simulation time (zoomed on the initial steps); and the final eigenvalue probability distribution.}
    \label{fig:150}
\end{figure}

\begin{figure}
    \centering
    \includegraphics[width=1\linewidth]{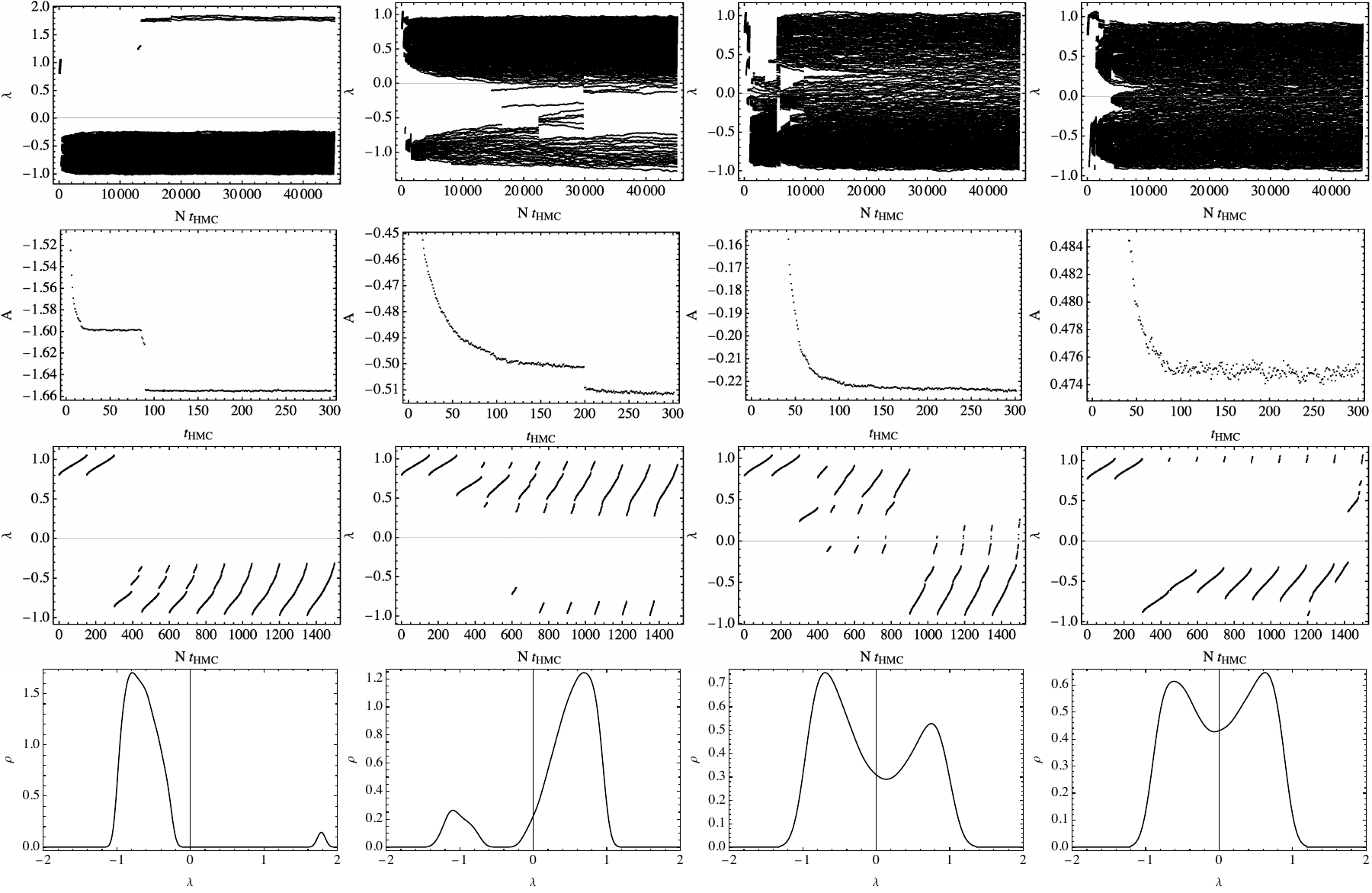}
    \caption{Results of the numerical simulations of the $N=150$ model specified in \eqref{action} with the cluster-algorithm improved HMC, columns are for $g=-4,-3.3,-3,-2$. The lines show: eigenvalues over simulation time; free energy over simulation time; eigenvalues over simulation time (zoomed on the initial steps); and the final eigenvalue probability distribution.}
    \label{fig:150C}
\end{figure}

The results are shown in Figures  \ref{fig:200}, \ref{fig:200C}, \ref{fig:comp}, \ref{fig:mig}, \ref{fig:100}, \ref{fig:100C}, \ref{fig:150}, \ref{fig:150C}, \ref{fig:FS}, \ref{fig:GW}, \ref{fig:c1c3}. We can see that, despite a limited number of HMC steps, the cluster algorithm found the correct solution. The simulations have not thermalised, but that is beyond the point; the goal was to show how the cluster algorithm outperforms the standard HMC in scanning the relevant parts of the configuration space; this is because the algorithm allows the simulation to make jumps, as shown by the discontinuities in the free energy plots. In Table \ref{comparison_table}, we see that the improved algorithm always converged to a state of lower free energy (in one case, both algorithms converged to the same free energy state within the precision).

For the simulations, we found the following values for the cluster algorithm to work best: $\sigma_\alpha=0.5,\mu_\beta=0.35,\sigma_\beta=0.3,\rho=0.3$. For these parameters, we tested different values between $0.2$ and $0.6$ with an increment of $0.05$ and chose the value that led to the most simulations ending in the correct vacuum (as predicted by the analytical description). This parameter optimisation is model-dependent, and when performing a simulation, one needs to perform a similar scan. For models with unknown solutions, the criterion can be the ability to find minimum free-energy states. On average, one cluster proposal was attempted per HMC sweep --- an aggressive procedure that can be turned off after thermalisation to save computational time.

\begin{table}[h] 
\centering
\begin{tabular}{|c|c|c|c|c|c|c|}
\hline
        & \multicolumn{2}{c|}{N=100} & \multicolumn{2}{c|}{N=150} & \multicolumn{2}{c|}{N=200} \\
\hline
g   & HMC & HMC$_C$ & HMC & HMC$_C$ & HMC & HMC$_C$ \\ \hline
-4   & -1.617 & \textbf{-1.669} & \textbf{-1.599} & \textbf{-1.599} & -1.329 & \textbf{-1.626} \\
-3.3  & -0.458 & \textbf{-0.540} & -0.440 & \textbf{-0.530} & -0.001 & \textbf{-0.504} \\
-3   & -0.052 & \textbf{-0.223} & -0.036 & \textbf{-0.219} & 0.504 & \textbf{-0.215} \\
-2   & 0.776 & \textbf{0.460} & 0.633 & \textbf{0.475} & 1.941 & \textbf{0.486} \\
\hline
\end{tabular}
\caption{Mean free energy of the model \eqref{action} over the final 25\% of simulation steps for the standard HMC and the cluster-algorithm-improved version (HMC$_C$). The lower of the two values in each column-pair is set in bold; HMC$_C$ reaches the lower state in all but one configuration – in that case, both methods reached the same free energy (within the precision).}
\label{comparison_table}
\end{table}

To show this, we ran a test with $N=100$ and $g=-4$. In Figures \ref{fig:100}, \ref{fig:100C}, one can observe that the system has two different possible vacua, one single-cut and one in which a small portion of eigenvalues moved to a distant new cut. We take the eigenvalues from the last step of the single-cut distribution from \ref{fig:100} and the position of the second peak from \ref{fig:100C}. We took one eigenvalue and, in 100 steps, manually moved it from one cut to the other; see Figure \ref{fig:mig}. We see that there is a peak in the free energy, and we compare the height of the peak, $\Delta_F A$, with the standard deviation of the free energy, $\delta_F A$, computed from the second half of the data shown in \ref{fig:100}. From the data, we obtained the estimate $\Delta_F A \approx 40 \times \delta_F A$, which means that fluctuating to the other solution is virtually impossible. However, this is an underestimate of the probability; one can choose a more aggressive step length (losing the acceptance rate), and the bulk can move spontaneously close to the origin to push the other eigenvalue by the repulsion, but still, it shows why the standard HMC can often remain in the false vacuum state. On the other hand, the cluster algorithm found the better solution in about 100 moves. The results also show why the eigenvalue-flipping algorithm fails to work in this case, as $\lambda \rightarrow -\lambda$ typically increases the free energy. 

To further demonstrate the algorithm's consistency and pace, we ran a simulation of $2000$ steps, resetting the system to the initial configuration after every $100$ steps. In Figure \ref{fig:comp}, we show the comparison of the standard HMC with the cluster algorithm HMC. We see that while the standard HMC is just beginning thermalisation, the cluster algorithm has already scanned different portions of the configuration space. 

\section{Conclusion}

In this paper, we presented a cluster algorithm that improves the standard HMC for matrix models with a rich vacuum structure. It was inspired by the eigenvalue-flipping algorithm of \cite{Kovacik:2022kfh} but is designed to work for models without the $\Phi \rightarrow -\Phi$ symmetry, where flipping alone fails to connect inequivalent local minima. The efficiency of the method relies on the fact that the behaviour of many matrix models is largely encoded in the eigenvalue distribution: by selecting a cluster of nearby eigenvalues and rescaling them coherently, the algorithm can propose moves that cross potential barriers which would be exponentially suppressed under standard local updates.

We demonstrated the method on the Dirac $(1,0)$ ensemble model \eqref{action}, which supports a variety of asymmetric multi-cut local minima for $g<-3.187$. Starting from a configuration close to a known false vacuum, the cluster algorithm reliably escapes within $O(100)$ moves for matrix sizes $N=100,\,150,\,200$, whereas standard HMC remains trapped on the timescales tested. The reason is quantified in Figure~\ref{fig:mig}: the barrier separating the false vacuum from the global minimum is $\Delta_F A \approx 40 \times \delta_F A$, where $\delta_F A$ is the typical fluctuation of the free energy --- spontaneous tunnelling is therefore effectively impossible in a standard simulation.

Each cluster move requires an eigenvalue decomposition, which scales as $O(N^3)$ and is more expensive than a local Metropolis step. The four parameters $(\sigma_\alpha,\mu_\beta,\sigma_\beta,p)$ require some tuning per model, and we provide empirical evidence of improved mixing rather than a proof of ergodicity, but for the class of models considered, the gain is substantial. We have chosen an update proposal that led to the Metropolis check \eqref{Metropolis}, one can choose a simpler version, for example, more symmetric, to simplify that expression. During our simulations, we tested removing the standard HMC update using the forces $-\partial S / \partial \Phi$ and observed that the $\hmcc$ algorithm still worked without it, indicating it found the free-energy minimum state. Therefore, it's plausible that an algorithm based solely on $\hmcc$ is a feasible choice, perhaps in cases where computing the forces is particularly difficult. 

Several directions are natural to pursue. For example, using a different procedure for moving clusters of eigenvalues, or even an adaptive method for the cluster algorithm, which might be adjusted during thermalisation. We have tested the method on various models, and the results are satisfying. The method is ready to be applied to more complex models. A natural extension is to coupled multi-matrix ensembles, where the notion of a cluster of eigenvalues must be generalised across the constituent matrices. More broadly, by enabling reliable sampling of the vacuum structure of matrix models with a nontrivial solution space, we hope this tool will be useful in the wider numerical programme for matrix models arising in M-theory, noncommutative geometry, and effective descriptions of quantum space. 

\acknowledgments This research was supported by VEGA 1/0025/23 \emph{Matrix models and quantum gravity} and VEGA 1/0604/26 \emph{Quantum structures of spacetime}. The authors would like to acknowledge the contribution of the COST Action CA23130, \textit{Bridging high and low energies in search of quantum gravity} and COST Action CA21109, \textit{Cartan geometry, Lie, Integrable Systems, quantum group Theories for Applications}. The authors would like to thank Dragan Prekrat for his comments. 

\FloatBarrier
\section{Appendix}
\subsection{Additional results for the Dirac $(1,0)$ model.}
Later, we present the remaining plots of the model \eqref{action} with the same parameters as for $N=200$, but now for $N=100$ and $N=150$.

\subsection{Example: The fuzzy sphere model}
\begin{figure}
    \centering
    \includegraphics[width=1\linewidth]{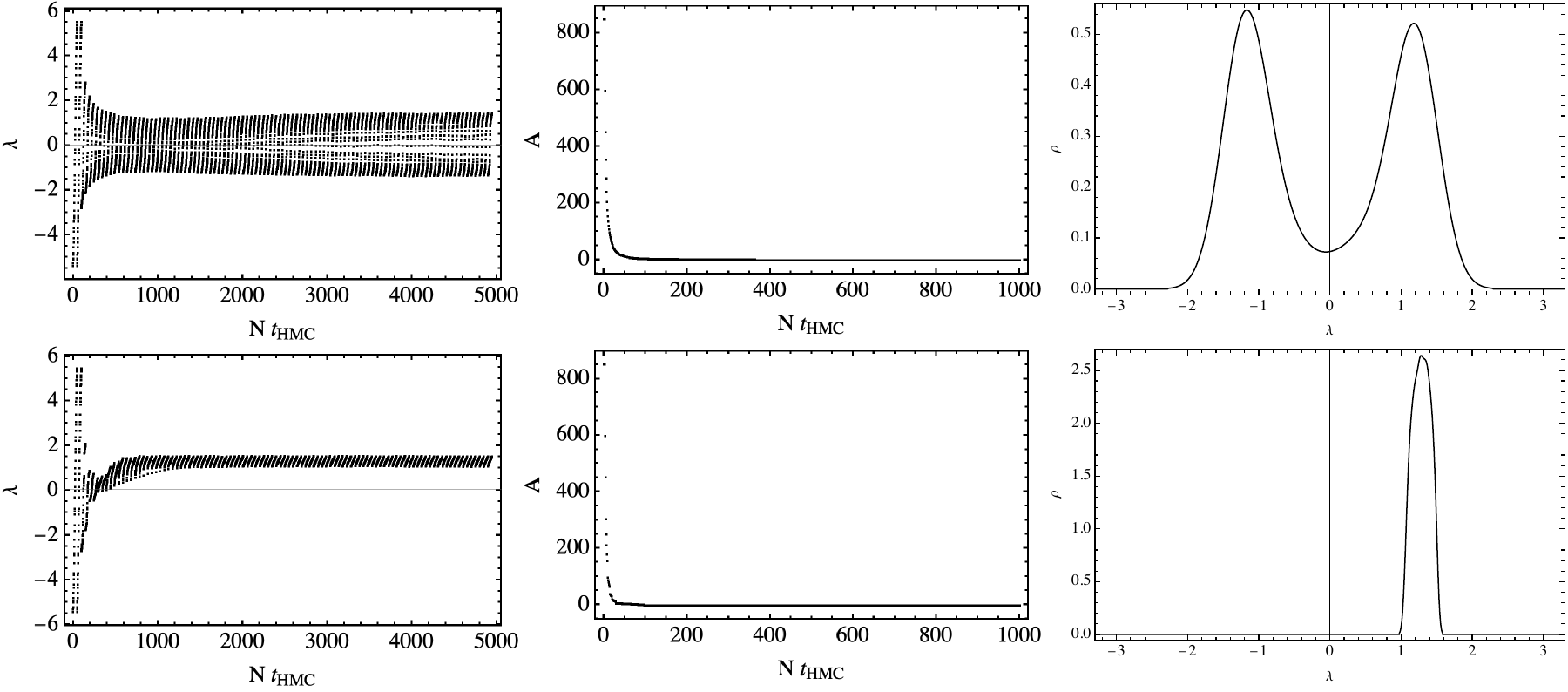}
    \caption{Comparison of $\mbox{HMC}$ and $\mbox{HMC}_C$ simulation of the fuzzy sphere model for $b=-1.2, c=0.04, N=49$. Plots on the left show the trajectories of eigenvalues, the middle ones show the evolution of the free energy, and those on the right show the eigenvalue distribution as reconstructed from the last $10$ simulations steps.  }
    \label{fig:FS}
\end{figure}

The fuzzy sphere model is defined by the action 
\begin{equation}
    S(\Phi) = N\mbox{Tr } \left( \Phi \Delta \Phi + b\Phi^2 + c \Phi^4 \right),
\end{equation}
where $\Delta$ is the Laplace operator defined by the double commutator of the size-$N$ representation of the $su(2)$ algebra, $\Phi$ are Hermitian matrices, see \cite{Kovacik:2018thy}. While the standard HMC algorithm found for the simulation of $b=-1.2, c=0.04, N=49$ preferable the symmetric two-cut phase, the $\mbox{HMC}_C$ algorithm found the asymmetric single-cut phase, the corresponding free energies were $F_{\mbox{\scriptsize{HMC}}}= -3.54$ and $F_{\mbox{\scriptsize{HMC}}_C}= -4.45$, results are shown in Figure \ref{fig:FS}. Simulations were initialised in a symmetric configuration. 

\subsection{Example: The Grosse-Wulkenhaar model}

\begin{figure}
    \centering
    \includegraphics[width=1\linewidth]{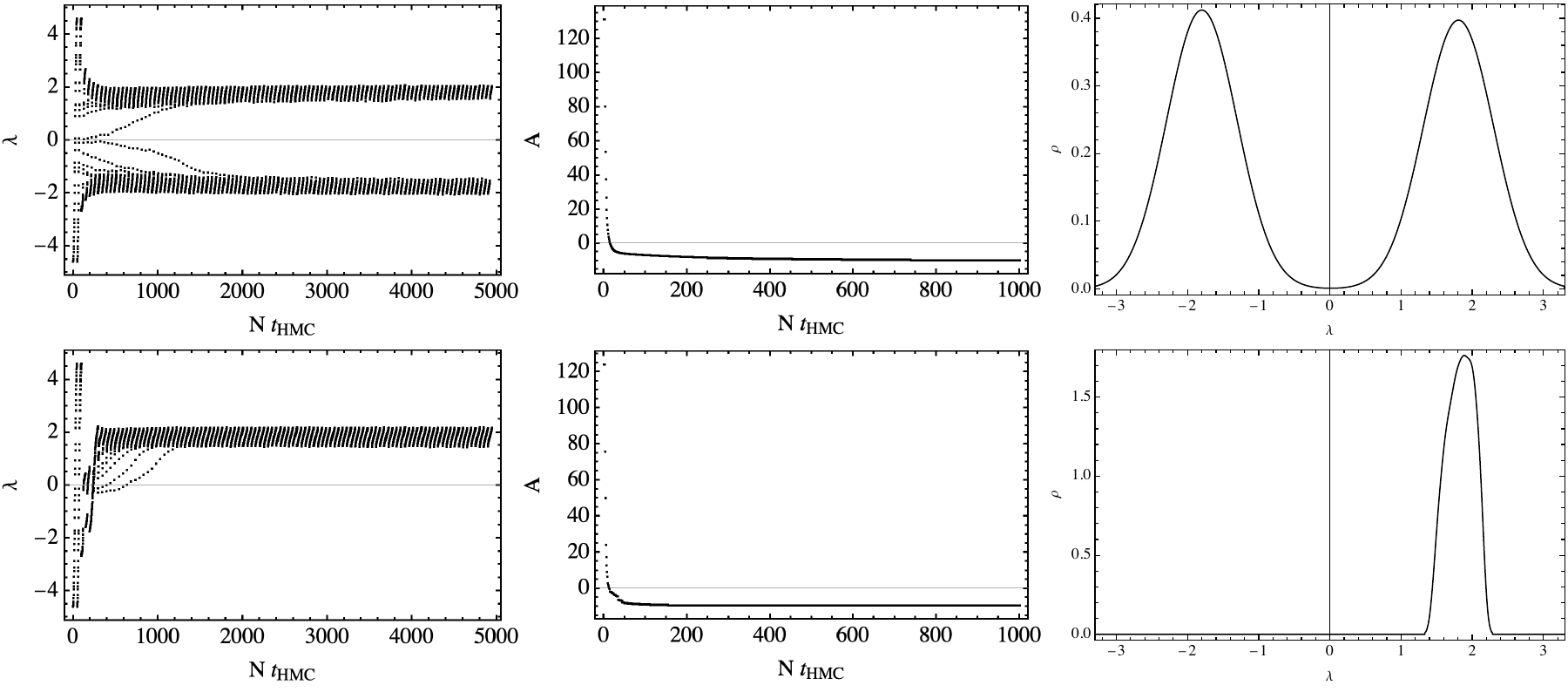}
    \caption{Comparison of $\mbox{HMC}$ and $\mbox{HMC}_C$ simulation of the Grosse-Wulkenhaar model for $g_r=0.01,g_2=7,g_4=1, N=49$. Plots on the left show the trajectories of eigenvalues, the middle ones show the evolution of the free energy, and those on the right show the eigenvalue distribution as reconstructed from the last $10$ simulations steps.  }
    \label{fig:GW}
\end{figure}

The Grosse-Wulkenhaar model is defined by the action 
\begin{equation}
    S(\Phi) = N  \mbox{Tr } \left( \Phi \Delta \Phi -g_r R \Phi^2- g_2 \Phi^2 + g_4 \Phi^4 \right),
\end{equation}
where $\Delta$ is the Laplace operator defined using rescaled Moyal plane coordinates, $\Phi$ are Hermitian matrices and $R=-\frac{16}{N} \mbox{diag }\left(1,2, \dots, N\right)$ is a constant matrix which captures the curvature of the space, we take its large-$N$ form, details can be found in \cite{Prekrat:2025viy}. We performed the simulations for $g_r=0.01,g_2=7,g_4=1, N=49$. We observed that the $\mbox{HMC}$ algorithm picked the symmetric two-cut phase while the $\mbox{HMC}_C$ found the asymmetric one-cut; their corresponding free energies were $F_{\mbox{\scriptsize{HMC}}}= -8.460$ and $F_{\mbox{\scriptsize{HMC}}_C}= -8.934$, results are shown in Figure \ref{fig:GW}. Simulations were initialised in a symmetric configuration. 

\subsection{Example: The asymmetric multi-trace model}

\begin{figure}
    \centering
    \includegraphics[width=1\linewidth]{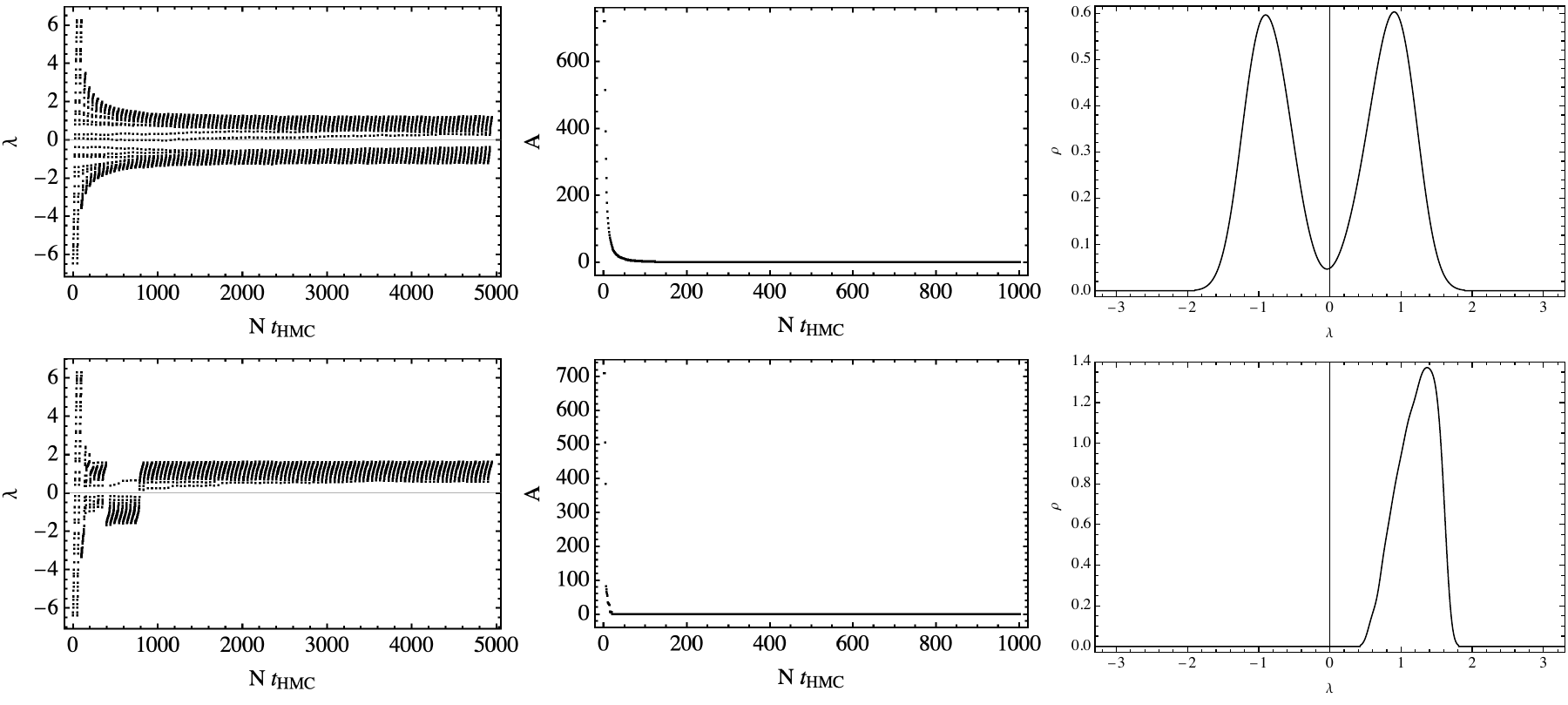}
    \caption{Comparison of $\mbox{HMC}$ and $\mbox{HMC}_C$ simulation of the asymmetric multi-trace model for $b=-2.6325, c=1.733, N=49$. Plots on the left show the trajectories of eigenvalues, the middle ones show the evolution of the free energy, and those on the right show the eigenvalue distribution as reconstructed from the last $10$ simulations steps.  }
    \label{fig:c1c3}
\end{figure}

The considered asymmetric multi-trace model was defined by the action
\begin{equation}
S(\Phi) = N \mbox{tr }\left(b\ \Phi^2 + c\ \Phi^4\right) - \mbox{tr } \Phi \mbox{ tr } \Phi^3.
\end{equation}
The model and its phase diagram are studied in detail in \cite{Bukor:2024kqy}. We used the values of $b=-2.6325, c=1.733, N=49$. For these parameters, the $\mbox{HMC}$ simulation found the symmetric two-cut phase while the $\mbox{HMC}_C$ simulation found the asymmetric one-cut phase (with a single eigenvalue left behind). The corresponding free energies were $F_{\mbox{\scriptsize{HMC}}}= -0.237$ and $F_{\mbox{\scriptsize{HMC}}_C}= -0.248$. Results are shown in Figure \ref{fig:c1c3}.

\FloatBarrier

\bibliographystyle{unsrt}
\bibliography{biblio}

\end{document}